\documentclass[aps,prc,twocolumn,floatfix,12pts,superscriptaddress]{revtex4}
\usepackage{epsfig}
\usepackage{amssymb}
\usepackage{amsmath}
\usepackage{color}
\usepackage{graphicx}
\usepackage{epsfig}
\usepackage{amsmath}
\usepackage{color}
\usepackage{graphicx}
\usepackage{multirow}
\usepackage{hyperref}
\usepackage{float}
\definecolor{blue}{rgb}{0.05, 0.05, 0.5}
\def \beq{\begin{equation}}
\def \eeq{\end{equation}}
\def \beqa{\begin{eqnarray}}
\def \eeqa{\end{eqnarray}}

\usepackage{xspace}

\begin{document}
\title{Initial state and evolution of hot and dense medium produced in isobaric
collisions at 200A GeV at RHIC}
\author{Amit Paul}
\email{a.paul@vecc.gov.in}

\author{Rupa Chatterjee}
\email{rupa@vecc.gov.in}
\affiliation{Variable Energy Cyclotron Centre, 1/AF, Bidhan Nagar, Kolkata-700064, India}
\affiliation{Homi Bhabha National Institute, Training School Complex, Anushaktinagar, Mumbai 400094, India}

\begin{abstract}
Isobaric collisions provide a unique opportunity to investigate how variations in the charge to mass ratio affect the final state observables produced in relativistic heavy ion collisions. Most importantly, isobaric systems that differ in their nuclear structure offer valuable insights into the underlying nuclear geometries, making them powerful tools to probe the role of nuclear structure using heavy ion collisions. We study the initial state and evolution of the hot and dense medium formed in Ru+Ru and Zr+Zr collisions at 200A GeV at RHIC using a relativistic hydrodynamical model. The initial geometry of the two isobaric collisions is found to influence the evolution of the hot and dense medium produced. The sensitivity of  photon production, charged particle spectra and anisotropic flow coefficients ($v_n$) to the initial geometry, including different orientations of the isobaric set have been studied in detail. Significant variations in anisotropic flow of photons and hadrons are observed, highlighting the role of  nuclear deformation in shaping final state observables. Moreover, photon anisotropic flow is found to be considerably more sensitive to the initial state than charged particle anisotropic flow, indicating that photon measurements in isobaric collisions have strong potential to constrain initial state modeling and improve our understanding of QGP properties in such systems.

\end{abstract}

\maketitle

\section{Introduction}

Collisions of heavy nuclei at top RHIC and LHC energies provide strong evidence for the formation of a Quark–Gluon Plasma (QGP)~\cite{qgp1,qgp2,qgp3,qgp4}. The observed large anisotropic flow of hadrons is successfully described within the framework of relativistic hydrodynamic models~\cite{v2_1, v2_2, kolb}. Isobaric collisions, involving nuclei with the same mass number and different proton numbers offer a unique and systematic framework to investigate the fundamental properties of QGP~\cite{J.Hammelmann:2020,G.Giacalone:2021,M.Adallah:2022,C.Zhang:2022,G.Nijs:2023,S.Zhao:2023,J.Wang:2024,J.Jia:2024}. The isobar run of Ru+Ru and Zr+Zr collisions at $\sqrt{s_{\text{NN}}} = 200$ GeV at RHIC was carried out with the aim of addressing several key physics objectives.

One of the initial motivations for studying isobaric collisions arises from the proposed search for the chiral magnetic effect (CME)~\cite{M.Adallah:2022}. Since the initial magnetic field scales with nuclear charge, Ru+Ru collisions are expected to generate stronger fields than Zr+Zr, making their comparison a useful way of investigating CME signals. 

The main advantage of isobaric collisions is the ability to compare final state observables between closely similar systems. Since the two colliding systems have the same mass number, observables such as charged particle multiplicity, initial energy  density of the produced systems as well as the hard process rates are expected to be nearly identical. These may reduce systematic uncertainties in the measured observables enabling a clearer separation of effects arising solely from differences in the initial geometry or nuclear structure.

 The difference in deformation parameters between the isobaric nuclei set Ru+Ru and Zr+Zr leads to distinct initial state geometries, especially when specific collision orientations (such as tip-tip or body-body) are selected. The $^{96}_{44}$Ru nucleus is known to possess a mild prolate deformation, whereas $^{96}_{40}$Zr has a triangular structure~\cite{G.Giacalone:2021,C.Zhang:2022,S.Zhao:2023,J.Wang:2024}. These geometric differences influence the initial spatial anisotropies ($\epsilon_2$, $\epsilon_3$, etc.), which in turn affect the anisotropic flow parameters.

Isobaric collisions are also expected to reveal how differences in nuclear geometry influence eccentricity distributions and other initial state parameters which play a crucial role in theoretical models used to describe experimental data.

We investigate the initial state as well as evolution of the hot and dense fireball produced in isobaric collisions at RHIC with the goal of qualitatively understanding their influence on final state observables. Charged particle and thermal photon spectra are computed for different orientations of the most central isobaric collisions. We study anisotropic flow under different initial conditions to see how photon observables provide additional insights beyond hadrons, helping to understand nuclear structure deformation and constrain the initial state produced in relativistic heavy ion collisions.

Direct photons are among the cleanest probes in relativistic heavy ion collisions, with their flow observables showing strong sensitivity to initial conditions.~\cite{dks, gabor, gale} Over the past two decades studies have shown that collision geometry, initial state fluctuations as well as initial parameters of theoretical model calculation significantly influence photon anisotropic flow making them a valuable tool for investigating early stage of QGP dynamics~\cite{rc_prl, gale1, rc1, monnai, zakharov, liu, pd, pd1}. Thus, measuring photon anisotropic flow in isobaric collisions can provide crucial insight and contribute to our understanding of the long standing direct photon puzzle~\cite{ratio}.

\section{Framework}
The evolution of the hot and dense matter produced in Ru+Ru and Zr+Zr collisions at 200A GeV at RHIC has been studied using a hydrodynamical model framework. The Glauber model formalism is employed to construct a smooth initial density distribution for the initial state.

A two-parameter Woods–Saxon nuclear density profile of the form is used~\cite{kolb, M.Miller:2007}:
\begin{equation}
    \rho(r,\theta) = \frac{\rho_0} {1 + \exp\left[\frac{r - R(\theta)}{\xi}\right]} 
\end{equation}
where, $\rho_0$ is the normalization factor, $\xi$ is surface diffuseness and $R(\theta)$ is the radius of the deformed nucleus which is given by
\begin{equation}
    R(\theta) = R_A \left[ 1 + \beta_2 Y_{20}(\theta) + \beta_3 Y_{30}(\theta) \right] \ .
\end{equation}
$R_A$ denotes the nuclear radius in the absence of deformation. The parameters $\beta_2$ and $\beta_3$ represent the quadrupole and octupole deformations respectively. The $Y_{20}(\theta)$ and $Y_{30}(\theta)$ are the spherical harmonics in the above equation.

Several slightly different values of the deformation parameters for Ru and Zr are reported in the literature~\cite{C.Zhang:2022,G.Nijs:2023,S.Zhao:2023,J.Wang:2024}. However, in this study we consider a single set of deformation parameters as listed in Table~\ref{tab:WS_parameters}, where ruthenium exhibits a prolate structure and zirconium displays a triangular deformation.

\begin{table}[H]
    \centering
    \renewcommand{\arraystretch}{1.1} 
    \begin{tabular}{|c|c|c|c|c|c|c|}
        \hline
        Nucleus & \begin{tabular}{c} \(A\) \end{tabular} & \begin{tabular}{c} \(R_A\) (fm) \end{tabular} & \begin{tabular}{c} \(\xi\) (fm) \end{tabular} & \begin{tabular}{c} \(\beta_2\) \end{tabular} & \begin{tabular}{c} \(\beta_3\) \end{tabular} \\
        \hline
        Ruthenium (Ru) & 96 & 5.09 & 0.46 & 0.162  & 0     \\
        Zirconium (Zr) & 96 & 5.02 & 0.52 & 0.06     & 0.20 \\
        \hline
    \end{tabular}    
    \caption{Nuclear and deformation parameters for Ru and Zr~\cite{C.Zhang:2022}.}
    \label{tab:WS_parameters}
\end{table}

It is important to note that the primary aim of this study is to demonstrate the qualitative differences in fireball evolution and final state observables arising from the deformed structures of the two isobaric systems. Therefore, small variations in the deformation parameter (in other set of parameters used for isobaric nuclei) are not expected to significantly alter the results.

We mainly consider two extreme orientations of the most central Ru+Ru and Zr+Zr collisions at RHIC, corresponding to the maximum and minimum initial spatial eccentricities namely the body–body and tip–tip configurations respectively. The tip–tip orientation leads to a smaller overlap region but higher particle multiplicity, while the body–body orientation results in the lowest multiplicity and the largest overlap region~\cite{uli_uu, uli_uu1}.
The fully overlapping collisions can be identified by examining the spectator energy deposition in the zero degree calorimeter. Even in the most central full overlap collisions  various intermediate orientations are  possible and results for these intermediate orientations are investigated in the later part of this study.

The charged particle multiplicity for both isobaric set has been estimated using the relation~\cite{B.Alver:2011}:

\begin{equation}
    \frac{dN_{\text{ch}}}{d\eta} = \frac{N_{\text{WN}}}{2} \left[ 0.78 \ln\left( \sqrt{s_{\text{NN}}} \right) - 0.4 \right] 
\end{equation}
here,  $N_{\rm WN}$ denotes the total number of wounded nucleons.

The initial entropy ($s$) and or energy density ($\varepsilon$) obtained from the relation: $\frac{dN_{\text{ch}}}{d\eta} \propto A_\text{T} \, s \, \tau$ and $ s \propto \varepsilon^{3/4}$. The transverse area is denoted as $A_T$ which is different for all configurations.

\begin{figure}
	\centerline{\includegraphics*[scale=0.35,clip=true]{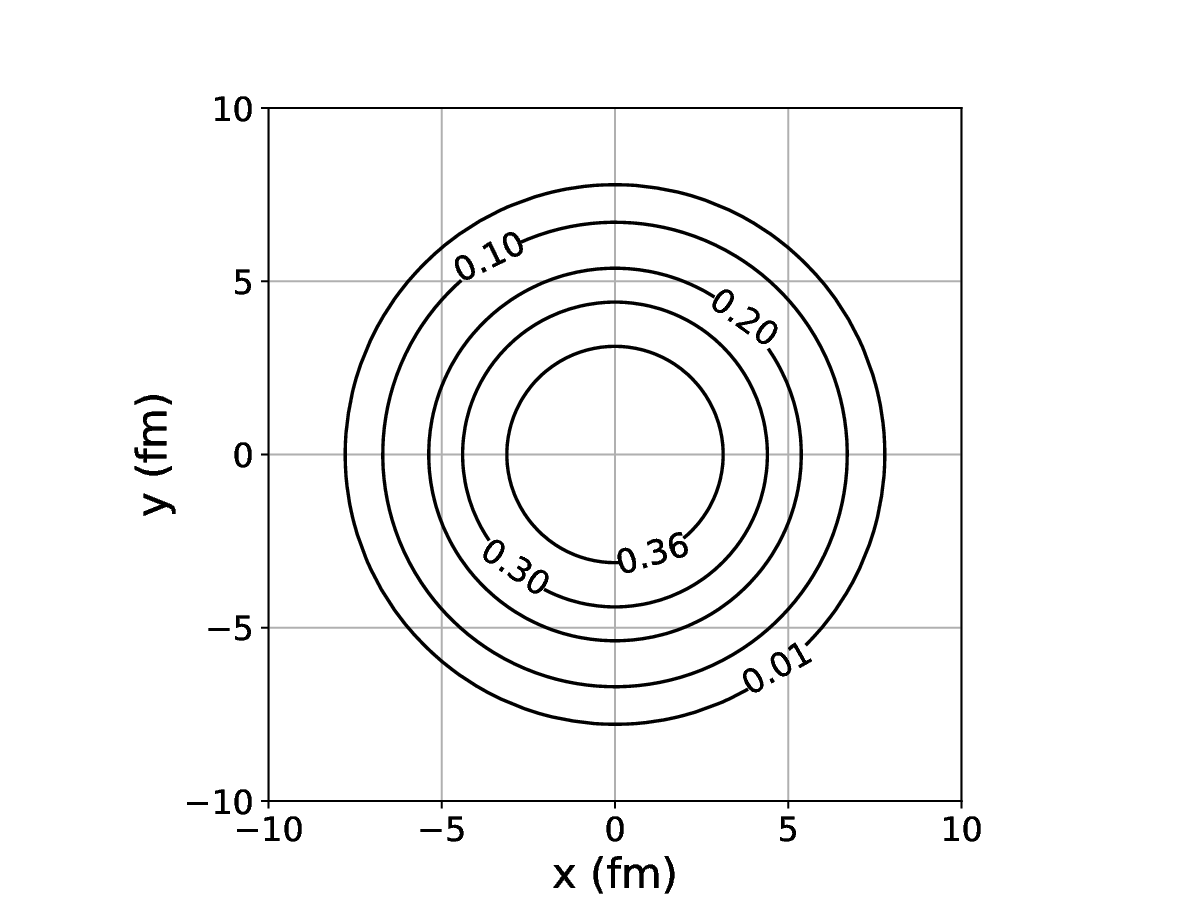}}
    \centerline{\includegraphics*[scale=0.35,clip=true]{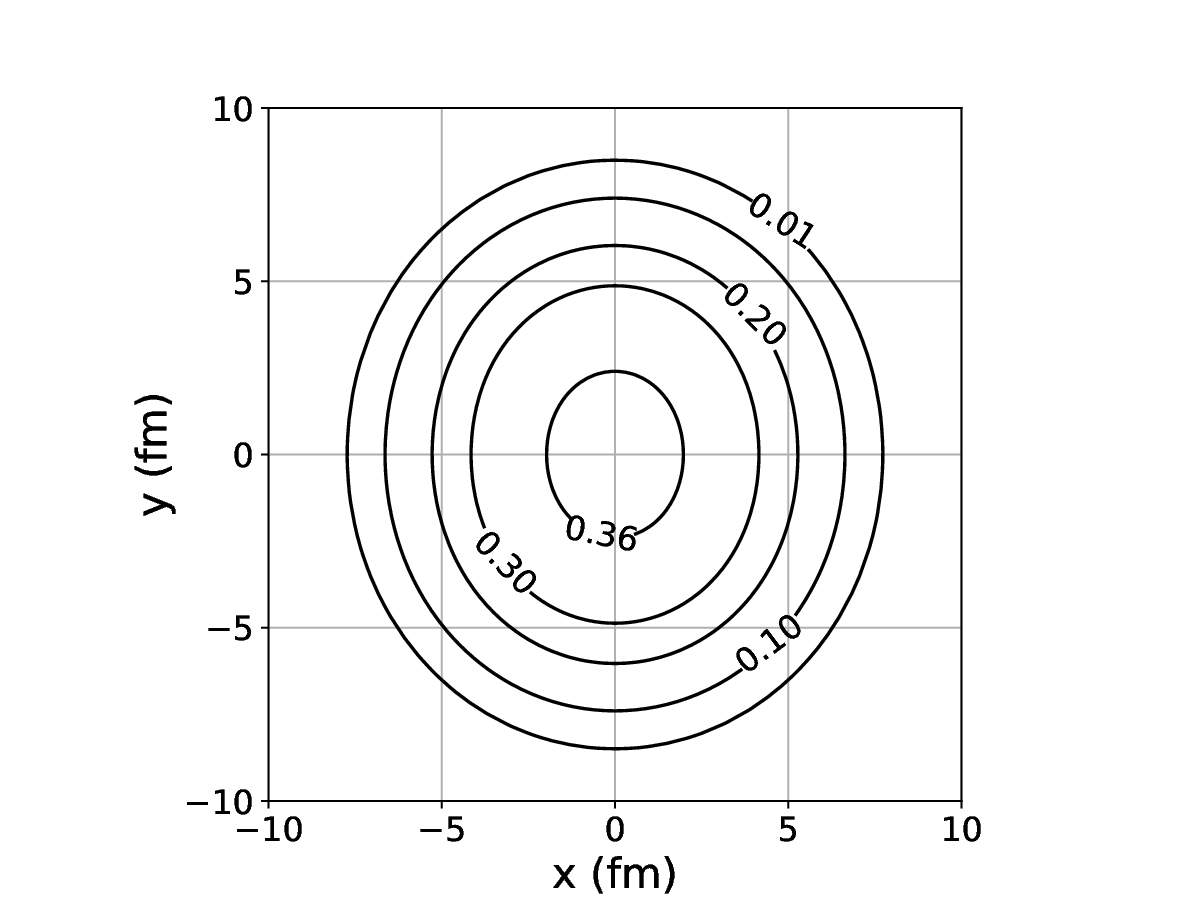}}

	\caption{Constant temperature contours (in GeV) for the two extreme cases of eccentricity in full overlapping Ru+Ru collisions at $\tau_0$ at 200A GeV at RHIC.}
	\label{fig.temp_contours_RuRu}
\end{figure}

\begin{figure}
	\centerline{\includegraphics*[scale=0.35,clip=true]{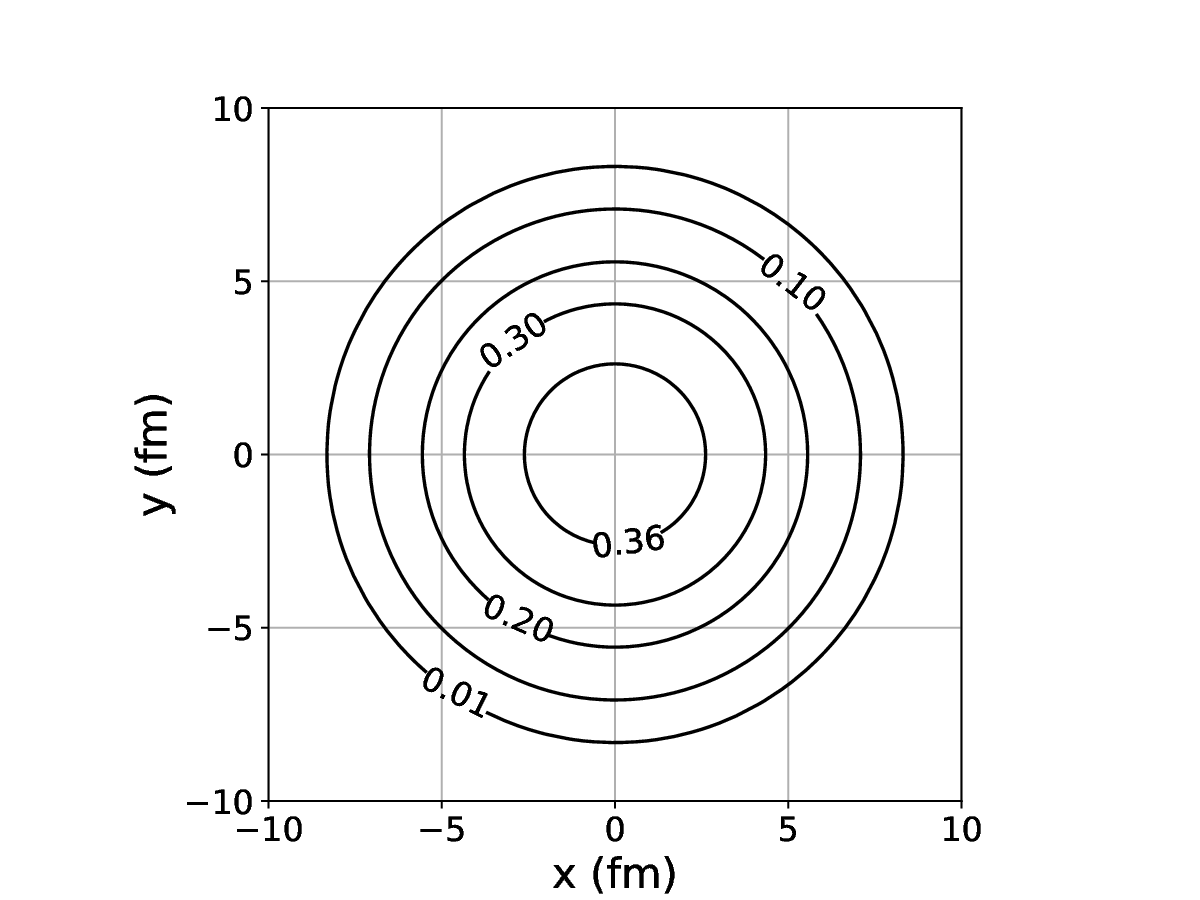}}
    \centerline{\includegraphics*[scale=0.35,clip=true]{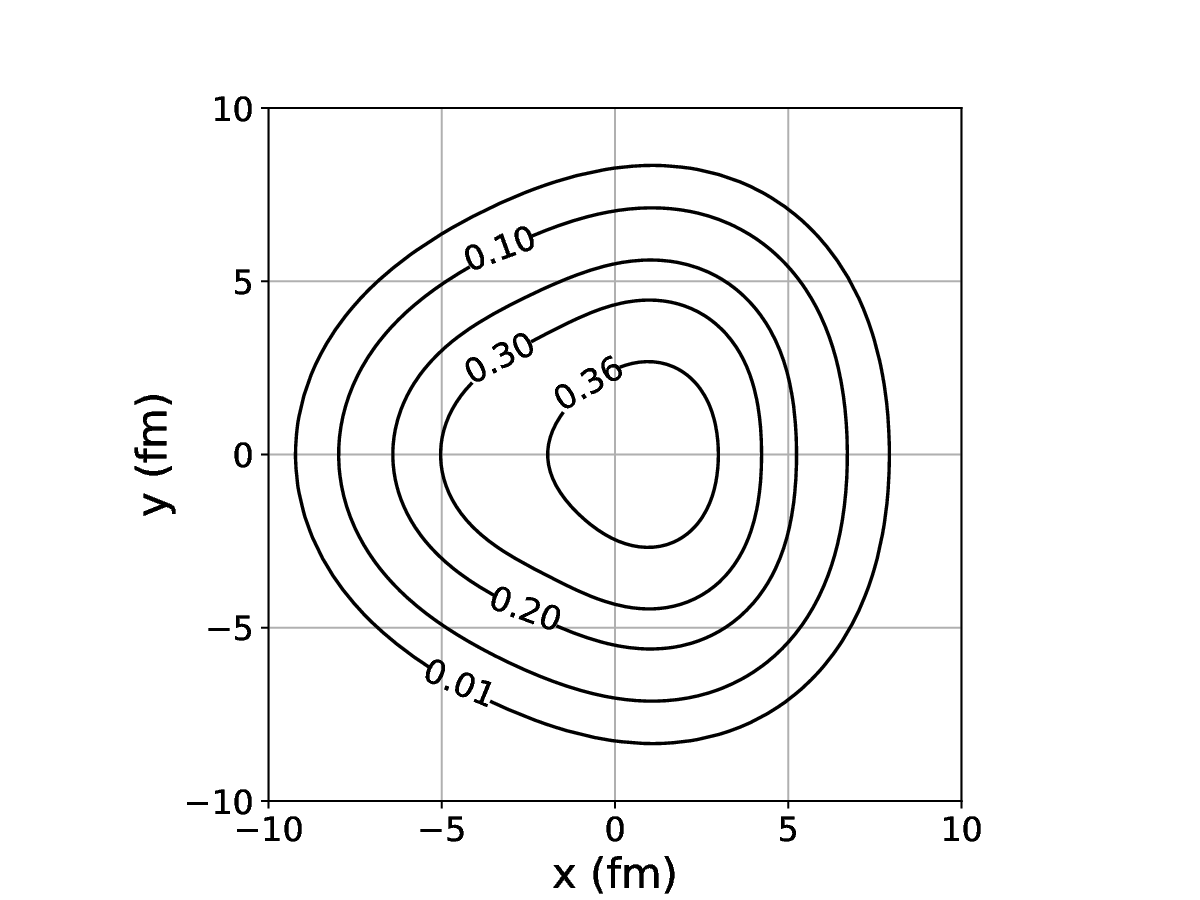}}

	\caption{Constant temperature contours (in GeV) for the two extreme cases of eccentricity in full overlapping Zr+Zr collisions at $\tau_0$ at 200A GeV at RHIC.}
	\label{fig.temp_contours_ZrZr}
\end{figure}
The shape of the initial energy density distribution in the transverse plane $(x,y)$ is parametrized as:
\begin{equation}
\varepsilon(x, y; b) = K \left[ (1 - \alpha)\, n_{\text{wn}}(x, y; b) + \alpha\, n_{\text{bc}}(x, y; b) \right]
\end{equation}
where $ n_{\text{wn}}(x, y; b)$ and  $n_{\text{bc}}(x, y; b)$ are the number of wounded nucleons and number of binary collisions per transverse area respectively at impact parameter $b$. The factor $\alpha$ is taken as 0.05. We have kept the same value of $\alpha$ that matches the particle spectra and multiplicity for 200A GeV Au+Au collisions at RHIC~\cite{B.Schenke:2010}.

$K$ in the expression above is obtained from the relations: $K=\varepsilon_0/\left[ (1 - \alpha)\, n_{\text{wn}}(0, 0; 0) + \alpha\, n_{\text{bc}}(0, 0; 0) \right]$ with $\varepsilon_0=\varepsilon(0,0;0)$.

The initial spatial eccentricities $\epsilon_n$ are obtain from the relation~\cite{B.Alver:2010}:
\begin{equation}
\epsilon_n = \frac{\sqrt{\langle r^2 \cos(n\phi) \rangle^2 + \langle r^2 \sin(n\phi) \rangle^2}}{\langle r^2 \rangle} \ .
\end{equation}

The MUSIC hydrodynamical framework is employed to study the evolution of the hot and dense matter produced in isobaric collisions at 200A GeV at midrapidity~\cite{B.Schenke:2010}. An initial formation time $\tau_0$ of 0.4 fm/$c$ is assumed for different orientations of Ru+Ru and Zr+Zr collisions at RHIC. Ideal hydrodynamical evolution is used to better isolate the effects of nuclear structure from other effects. A lattice based equation of state is considered  with a final freeze out temperature of 137 MeV for all cases~\cite{P.Huovinen:2010}. The nucleon–nucleon inelastic cross section ($\sigma_{\rm NN}$) is taken as 42 mb and the corresponding central energy densities are listed in Table~\ref{tab:hydro_parameters}. The total charged particle multiplicity varies in the range of 320 to 338 for the most central set of  isobaric collisions at 200A GeV at RHIC.

The Cooper-Frye prescription is used to calculate the charged particle spectra from the freeze-out hyper surface~\cite{Cooper:1974} and $p_T$ spectra are found to give  a reasonable agreement of the preliminary STAR experimental data (shown in the result section).

The production of thermal photons from different collision geometry is estimated by integrating the thermal photon emission rates (i.e. $R=E\frac{dN}{d^3pd^4x}$) over entire space-time evolution:
		\begin{eqnarray}
	{E \frac{dN}{d^3p} \ =  \int \it{R}\big(E^*(x),T(x)\big)d^4x} \ .
	\label{dn_phot}
\end{eqnarray}
The $T(x)$ in the above equation is the local temperature and $E^*(x)=p^\mu u_\mu(x)$, where $p^\mu$ represents the photon 4-momentum and $u_\mu$ is the local 4-velocity of the flow field. We used the complete leading-order emission rates from Ref.~\cite{Arnold:2001ms}  to evaluate photon production from the QGP medium and the parameterized rates from Ref.~\cite{Turbide:2003si}  for the hadronic phase. \\

\begin{table}[H]
    \centering
    \renewcommand{\arraystretch}{1.2} 
    \begin{tabular}{|p{3.cm}|c|c|c|c|}
        \hline
        $AA$ & $\varepsilon_0$ & $\epsilon_2(\tau_0)$ & $\epsilon_3(\tau_0)$ & $\langle T(\tau_0)\rangle$ \\
        & (GeV/fm$^3$) &  &  & (MeV) \\
        \hline
        \raggedright Ru+Ru tip-tip & 43.28 & 0 & 0 & 350 \\
        \raggedright Ru+Ru body-body & 34.79 & 0.146 & 0 & 333 \\
        \raggedright Zr+Zr tip-tip & 38.02 & 0 & 0 & 337 \\
        \raggedright Zr+Zr body-body & 37.21 & 0.062 & 0.158 & 336 \\
        \hline
    \end{tabular}
    \caption{Initial parameters from 200A GeV Ru+Ru and Zr+Zr collisions at RHIC.}
    \label{tab:hydro_parameters}
\end{table}

\section{Results and discussions}
The constant temperature contours for fully overlapping Ru+Ru collisions at 200A GeV at RHIC at time $\tau_0$ are shown in Fig.~\ref{fig.temp_contours_RuRu} illustrating two extreme cases of initial eccentricity. Due to the mildly prolate shape of the ruthenium nuclei, the initial overlap region deviates from a perfect circular geometry for body-body configuration [lower panel]. In contrast, the tip-tip configuration results in nearly concentric circular contours of constant temperature [upper panel] .

The constant temperature contours for the tip–tip configuration of Zr+Zr collisions [upper panel of Fig.~\ref{fig.temp_contours_ZrZr}] show a qualitatively similar behavior to those in Ru+Ru tip–tip collisions  characterized by nearly concentric circular contours arising from the symmetric overlap geometry. However, in the body-body configuration of Zr+Zr collisions the initial overlap region displays a significant triangular deformation and pronounced spatial anisotropy arising from the intrinsic deformation. 

Such distinct initial configurations [Figs.~\ref{fig.temp_contours_RuRu} and ~\ref{fig.temp_contours_ZrZr}] are expected to influence the evolution of the quark gluon plasma and leave clear imprints on final state observables, particularly in the anisotropic flow coefficients.

\begin{figure}
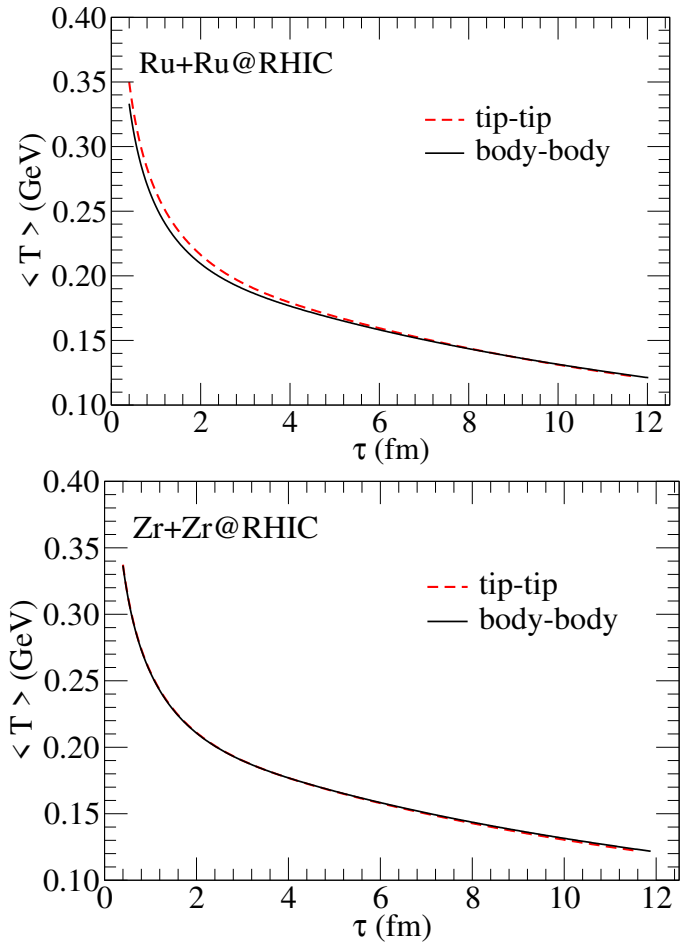

	\centerline{\includegraphics*[scale=0.35,clip=true]{ru_temp.eps}}
    \centerline{\includegraphics*[scale=0.35,clip=true]{zr_temp.eps}}

	\caption{(Color online)  Time evolution of average temperature $\langle T \rangle$ for tip-tip and body-body orientations of Ru+Ru (upper panel) and Zr+Zr (lower panel) collisions at 200A GeV at RHIC.}
	\label{fig.temp_evolution}
\end{figure}

\begin{figure}
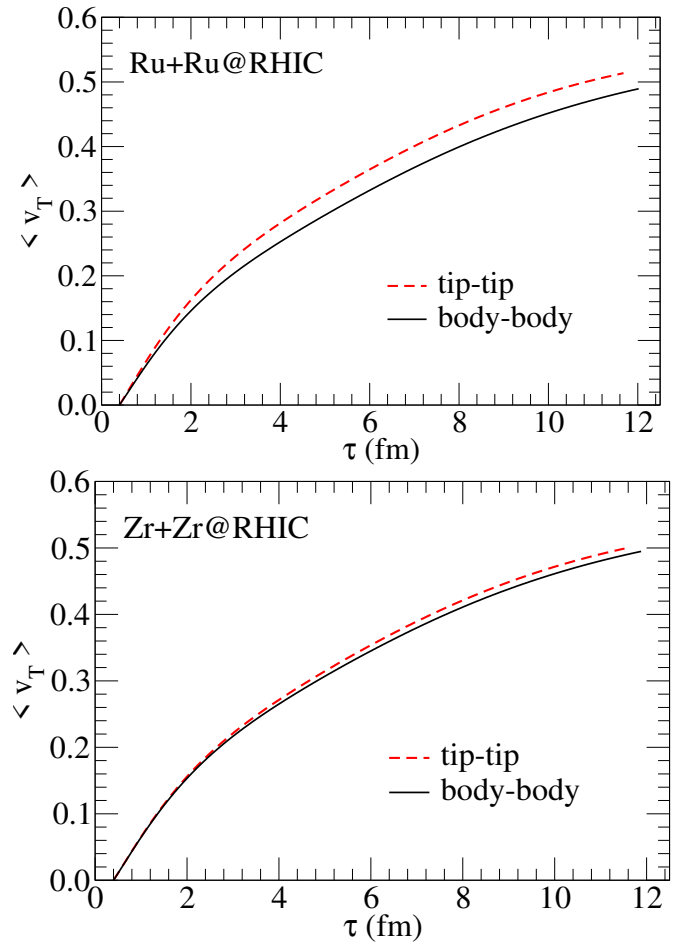

	\centerline{\includegraphics*[scale=0.35,clip=true]{ru_vt.eps}}
    \centerline{\includegraphics*[scale=0.35,clip=true]{zr_vt.eps}}

	\caption{(Color online) Time evolution of average transverse flow velocity $\langle v_T \rangle$ for tip-tip and body-body orientations of Ru+Ru (upper panel) and Zr+Zr (lower panel) collisions at 200A GeV at RHIC.}
	\label{fig.vt_evolution}
\end{figure}

Figs~\ref{fig.temp_evolution} and~\ref{fig.vt_evolution} present the time evolution of the average temperature $\langle T \rangle$ and transverse flow velocity $\langle v_T \rangle$ for the two extreme cases of initial eccentricity, obtained from hydrodynamic simulations. The average of a quantity $f(\tau)$ at a particular $\tau$ is obtained using the relation: 
\begin{equation}
\langle f (\tau) \rangle = \frac{\int \int dx \, dy \, \varepsilon(x,y,\tau) \, f(x,y,\tau)}{\int \int dx \, dy \, \varepsilon(x,y,\tau)} \ .
\end{equation}
The energy density on the transverse plane $\varepsilon(x,y,\tau)$ is used as a weight factor to estimate the average value.
For both Ru+Ru and Zr+Zr collisions, the initial average temperature lies in the range of 330 to 350 MeV. In the case of Ru+Ru collisions $\langle T \rangle$ exhibits a slight difference between the tip-tip and body-body configurations. In contrast, for Zr+Zr collisions the evolution of $\langle T \rangle$ remains nearly similar for both configurations indicating a minimal sensitivity to the initial geometry.

The average transverse flow velocity $\langle v_T \rangle$ however, shows a noticeable difference between the two configurations in Ru+Ru collisions. In the tip-tip configuration, a higher energy density is deposited in a relatively smaller overlap volume leading to a more rapid build-up of transverse flow. As a result, the rise in $\langle v_T \rangle$ is sharper compared to the body-body case. Although $\langle v_T \rangle$ is marginally larger for tip-tip case compared to body-body of Zr+Zr collisions, the difference between the two configurations is not significant.

It is important to note that the initial elliptic eccentricity $\epsilon_2$ significantly influences the development of transverse flow during the medium evolution. While the triangular eccentricity $\epsilon_3$ in Zr+Zr collisions is found to be larger than the elliptic eccentricity $\epsilon_2$ in Ru+Ru, its impact on the build-up of average transverse flow is relatively less.

\subsection{Hadron spectra and anisotropic flow}
\begin{figure}
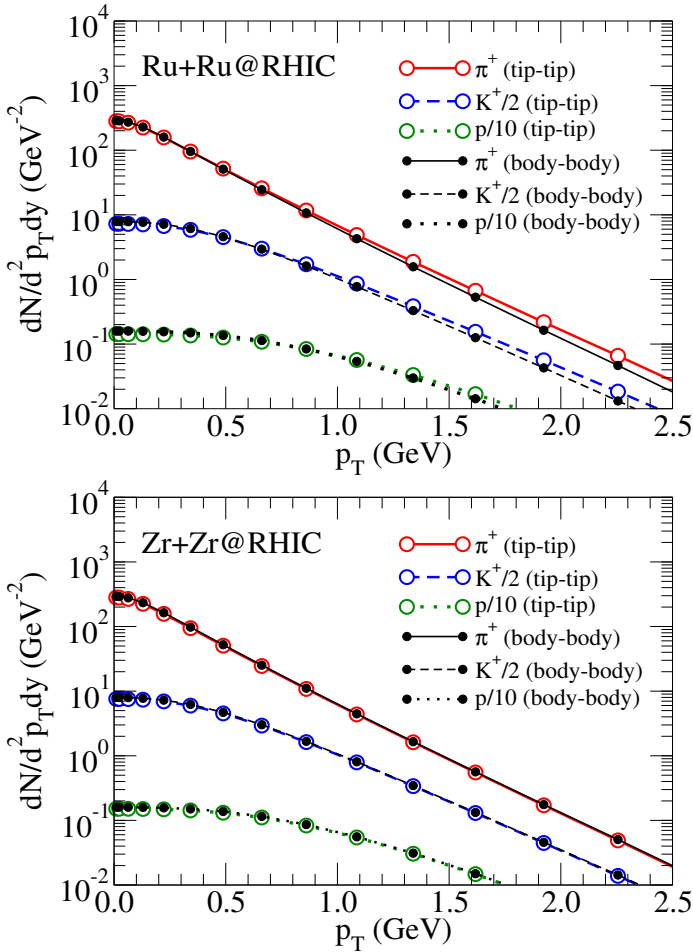

	\centerline{\includegraphics*[scale=0.35,clip=true]{ru_pt.eps}}
    \centerline{\includegraphics*[scale=0.35,clip=true]{zr_pt.eps}}

	\caption{(Color online) Charged hadron spectra for tip-tip and body-body orientations of Ru+Ru (upper panel) and Zr+Zr (lower panel) at 200A GeV at RHIC.}
	\label{fig.pt_spectra}
\end{figure}


\begin{figure}
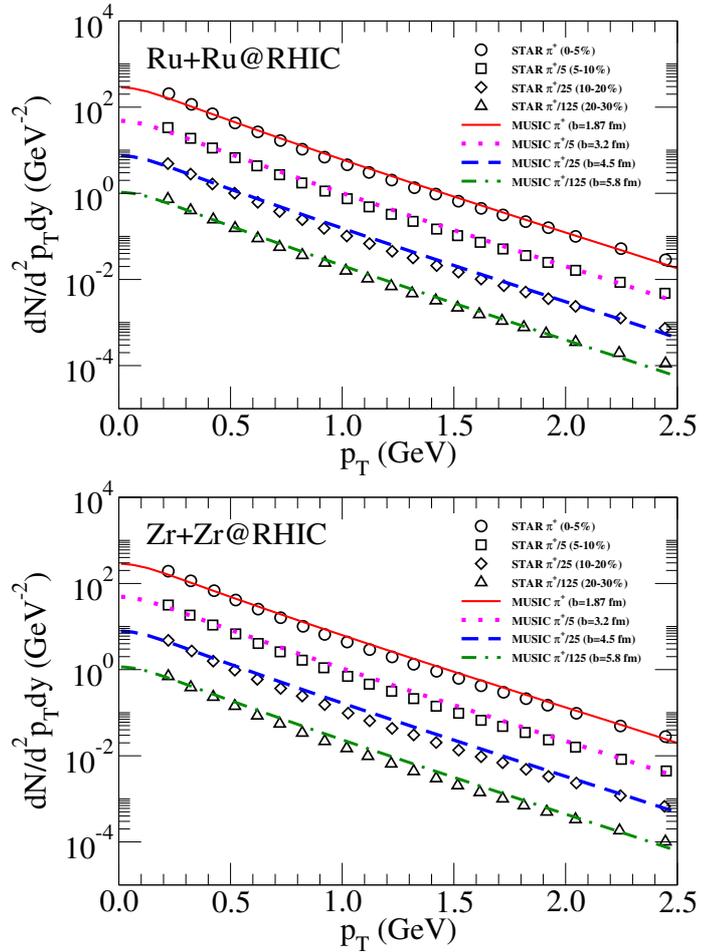

	\centerline{\includegraphics*[scale=0.35,clip=true]{cent_ru.eps}}
    \centerline{\includegraphics*[scale=0.35,clip=true]{cent_zr.eps}}

	\caption{(Color online) Centrality dependent charged pion $p_T$ spectra along with STAR preliminary data for 200A GeV Ru+Ru [upper panel] and Zr+Zr [lower panel] collisions at RHIC.}
	\label{fig.centrality_spectra}
\end{figure}

The transverse momentum ($p_T$) spectra of charged pions, kaons and protons for the two isobaric collision systems are presented in Fig.~\ref{fig.pt_spectra}. The results correspond to the tip-tip and body-body configurations of the most central Ru+Ru and Zr+Zr collisions at 200A GeV at RHIC. In the case of Ru+Ru collisions, the slightly higher average transverse flow velocity and initial temperature in the tip-tip configuration lead to minor differences in the charged particle spectra particularly at larger $p_T$ ($ > 1$ GeV) region. However, these variations are expected to lie within experimental uncertainties. In contrast, for Zr+Zr collisions, the $p_T$ spectra for the two configurations are nearly identical reflecting the similar hydrodynamic evolution resulting from comparable initial conditions.

We have checked that the charged pion $p_T$ spectra from our calculation provide a reasonable agreement with the STAR  preliminary experimental data~\cite{STAR_preliminary} for both Ru+Ru and Zr+Zr collisions at different centrality bins as shown in Fig.~\ref{fig.centrality_spectra}.

\begin{figure}
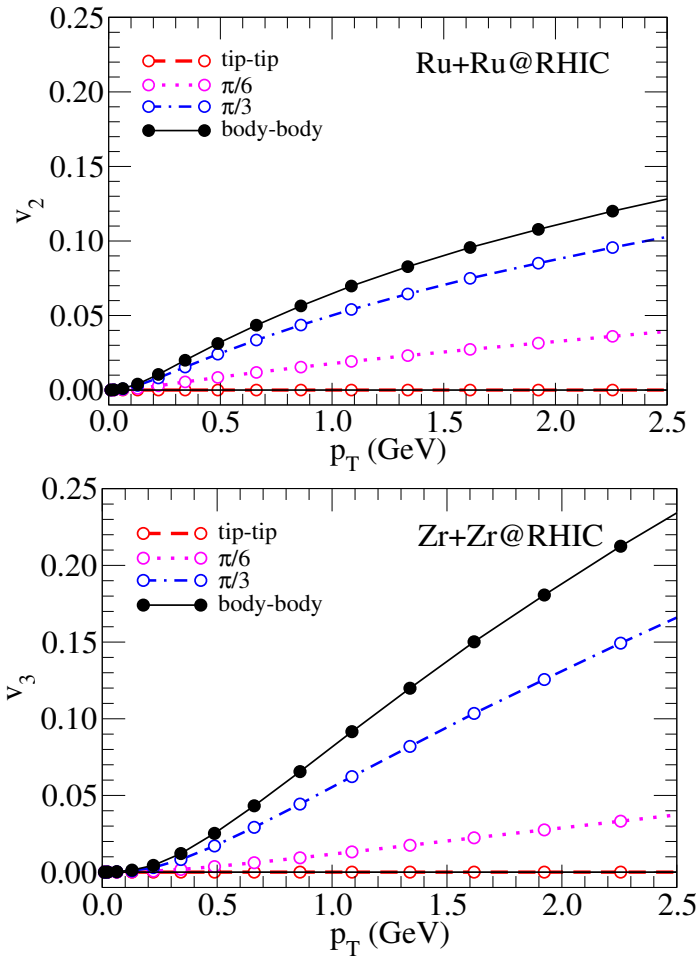

	\centerline{\includegraphics*[scale=0.35,clip=true]{ru_v2pi.eps}}
    \centerline{\includegraphics*[scale=0.35,clip=true]{zr_v3pi.eps}}

	\caption{(Color online) Elliptic and triangular flow of pions from most central Ru+Ru (upper panel) and Zr+Zr (lower panel) collisions considering $\tau_0=0.4$ fm/$c$ at 200A GeV at RHIC.}
	\label{fig.pion_flow1}
\end{figure}

The elliptic and triangular flow coefficients ($v_2$ and $v_3$) of charged pions for Ru+Ru and Zr+Zr collisions are shown in Fig.~\ref{fig.pion_flow1}. As expected, the elliptic flow parameter $v_2(p_T)$ increases with $p_T$ and reaches a maximum for the body-body configuration of Ru+Ru collisions. Notably, even a mild prolate deformation in the Ru nucleus leads to a significantly large elliptic flow due to the resulting anisotropic initial geometry.

In contrast, the tip-tip Ru+Ru configuration shows a nearly vanishing $v_2$ as it corresponds to an almost azimuthally symmetric initial overlap region with minimal eccentricity and fluctuations. Intermediate configurations with orientation angles of $\pi/6$ and $\pi/3$ between tip-tip and body-body produce finite values of $v_2$, reflecting a gradual buildup of initial eccentricity.

In Zr+Zr (body-body) collisions the magnitude of the triangular flow is found to be larger than the elliptic flow observed in Ru+Ru collisions. This is attributed to a larger initial triangular eccentricity $\epsilon_3$ in the Zr+Zr system compared to the elliptic eccentricity $\epsilon_2$ in Ru+Ru.

\begin{figure}
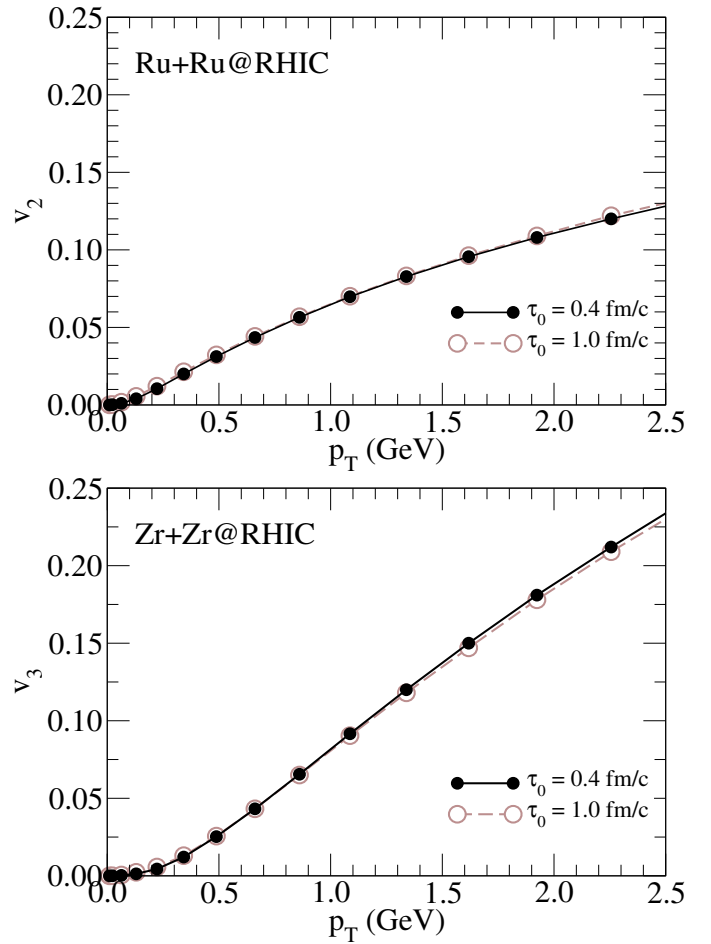

	\centerline{\includegraphics*[scale=0.35,clip=true]{dfrt_tau_v2ru.eps}}
    \centerline{\includegraphics*[scale=0.35,clip=true]{dfrt_tau_v3zr.eps}}

	\caption{(Color online) Anisotropic flow of pions from body-body orientations of  Ru+Ru (upper panel) and Zr+Zr (lower panel) for $\tau_0  = 0.4$  and $1.0$ fm/$c$  at 200A GeV at RHIC.}
    \label{fig.pion_flow2}
\end{figure}

The anisotropic flow of pions for two different initial formation time is shown in Fig.~\ref{fig.pion_flow2}. The value of $\tau_0$ is increased from 0.4 fm/$c$ to 1.0 fm/$c$ by considering  the total entropy and the charged particle rapidity density fixed. The anisotropic flow parameters are found to depend on the formation time only marginally for pions. 
 
\subsection{Thermal photon spectra and anisotropic flow}
\begin{figure}
	\centerline{\includegraphics*[scale=0.35,clip=true]{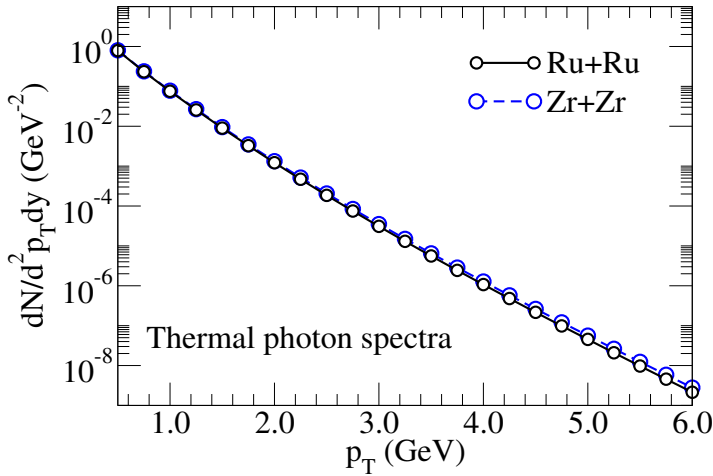}}

	\caption{(Color online) Thermal photons spectra from 200A GeV Ru+Ru and Zr+Zr collisions at RHIC considering $\tau_0=0.4$ fm/$c$.}
	\label{fig.photon_spectra}
\end{figure}

\begin{figure}
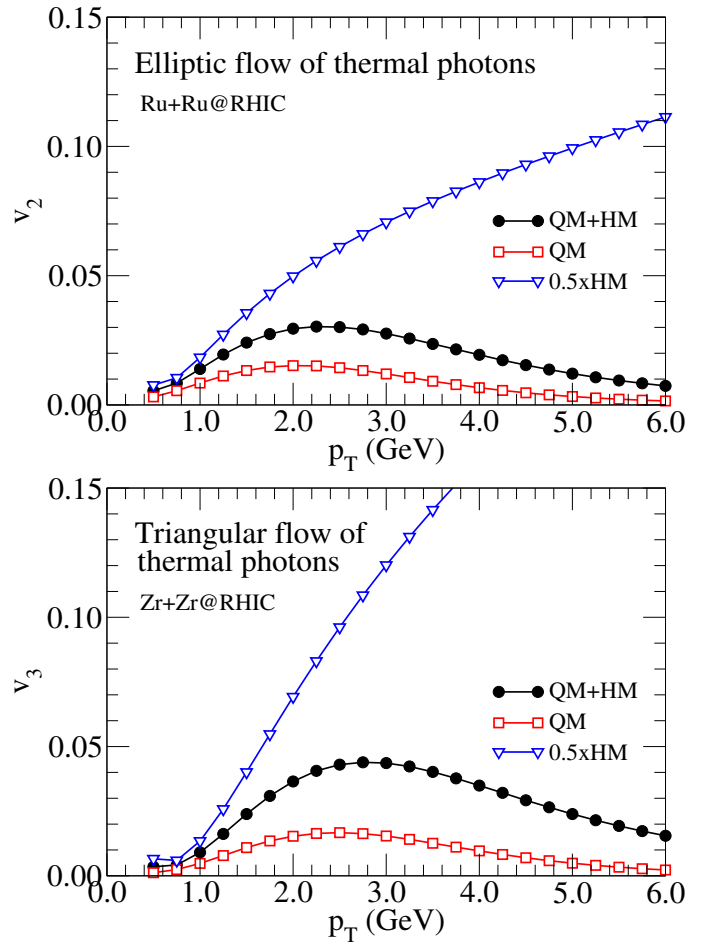

	\centerline{\includegraphics*[scale=0.35,clip=true]{ruv2ph.eps}}
    \centerline{\includegraphics*[scale=0.35,clip=true]{zrv3ph.eps}}

	\caption{(Color online) Elliptic (upper panel) and triangular flow (lower panel) of thermal photons from most central Ru+Ru and Zr+Zr collisions respectively for $\tau_0  = 0.4$ fm/$c$  at 200A GeV at RHIC along with individual contributions from QGP (QM) and hadronic matter (HM) to the total photon $v_n$(QM+HM). }
	\label{fig.photon_flow_q&h}
\end{figure}

\begin{figure}
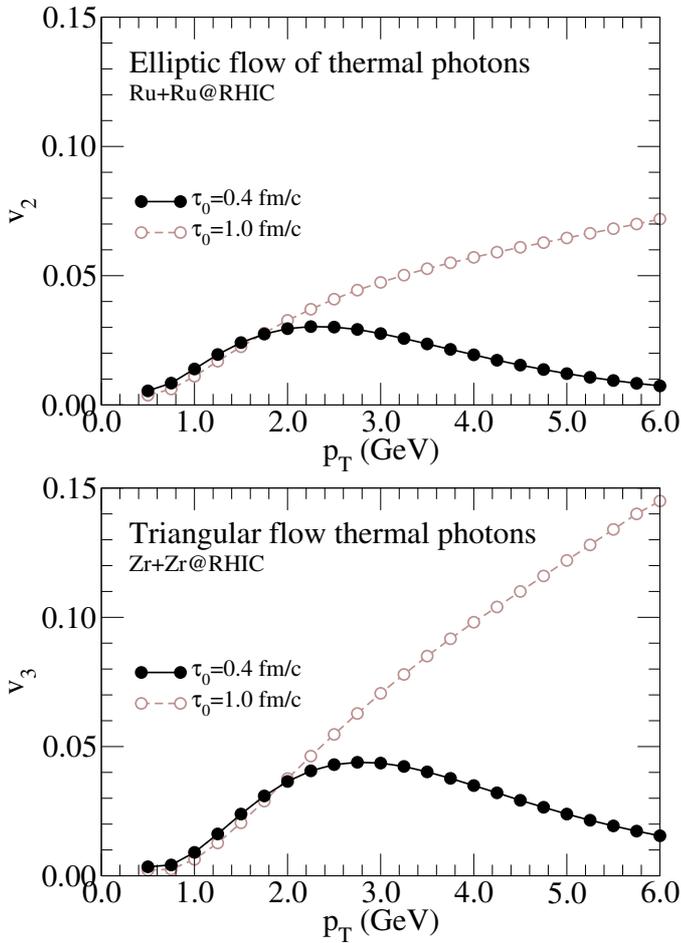

	\centerline{\includegraphics*[scale=0.35,clip=true]{ru_v2ph.eps}}
    \centerline{\includegraphics*[scale=0.35,clip=true]{zr_v3ph.eps}}

	\caption{(Color online) Elliptic (upper panel) and triangular flow (lower panel) of thermal photons from most central Ru+Ru and Zr+Zr collisions respectively for $\tau_0  = 0.4$  and $1.0$ fm/$c$  at 200A GeV at RHIC.}
	\label{fig.photon_flow}
\end{figure}

The spectra and anisotropic flow of thermal photons are presented in Figs.~\ref{fig.photon_spectra},~\ref{fig.photon_flow_q&h} and~\ref{fig.photon_flow}. The photon spectra for body-body configurations of both Ru+Ru and Zr+Zr collisions at 200A GeV at RHIC are shown as functions of $p_T$. Similar to charged particles, the thermal photon spectra show little sensitivity to the orientation angle of the colliding nuclei or to the structural differences between the two isobaric systems.

However, the anisotropic flow of photons shows strong sensitivity to the initial geometry as expected. The triangular flow $v_3$ in body-body Zr+Zr collisions is found to be significantly larger than the elliptic flow $v_2$ observed in Ru+Ru collisions for similar configurations. Additionally, the peaks of the thermal photon $v_2$ (for Ru+Ru) and $v_3$  (for Zr+Zr) distributions occur at different $p_T$ values, with the $v_3$ peak shifted to slightly higher $p_T$ compared to the $v_2$ peak for $\tau_0$ = 0.4 fm/$c$. 

The relative contributions of the plasma (QM) and hadronic matter (HM) phases determine the position of the peak as shown in Fig.~\ref{fig.photon_flow_q&h}~\cite{rc_prl}. The total photon $v_n$(QM+HM)  tend to follow the nature of $v_n$(QM) with increase in transverse momentum values. 
When both the QM and HM contributions are added with appropriate weight factor (photon yield from respective phases), the total $v_n$ is observed to be  larger than $v_n$(QM) but much smaller than $v_n$(HM) reflecting the smaller number of photons emitted from the hadronic phase at larger $p_T$ values. The relative contribution of the $v_n$(HM) is found to be larger than $v_n$(QM) for $v_3$ than for $v_2$ which results in the shift of the peak position at a relatively larger $p_T$ value for  the photon $v_3$. 

The anisotropic flow of thermal photons is found to be highly sensitive to the choice of $\tau_0$, in contrast to hadrons~\cite{tau_1, tau2} [see Fig.~\ref{fig.photon_flow}]. A larger $\tau_0$ leads to an increase in $v_n$ particularly in the high $p_T$ region, since the relative contribution from the QGP phase decreases with increasing $\tau_0$.

\begin{figure}
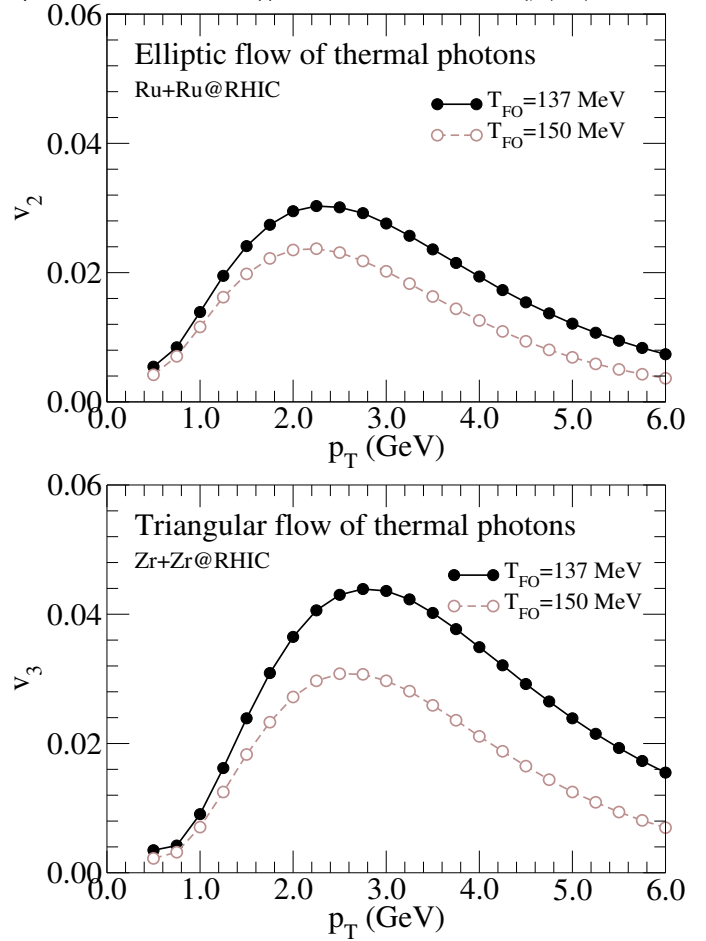

	\centerline{\includegraphics*[scale=0.35,clip=true]{dfrt_fo_v2ru.eps}}
    \centerline{\includegraphics*[scale=0.35,clip=true]{dfrt_fo_v3zr.eps}}

	\caption{(Color online) Elliptic (upper panel) and triangular flow (lower panel) of thermal photons from most central Ru+Ru and Zr+Zr collisions at 200A GeV at RHIC at $T_{\rm FO}$=137 and 150 MeV at fixed $\tau_0=0.4$ fm/$c$.}
	\label{fig.photon_flow2}
\end{figure}

The photon anisotropic flow also shows a strong dependence on the freeze-out condition. For a fixed $\tau_0$, a 10\% increase in $T_{\rm FO}$ (from 137 MeV to 150 MeV) leads to a significant reduction in $v_n$ as shown in Fig.~\ref{fig.photon_flow2}. This is because a larger $T_{\rm FO}$ suppresses the hadronic phase contribution which typically carries a higher $v_n$ compared to the QGP phase alone.

Furthermore, the contribution from prompt photons is expected to be similar for both collision systems and orientation configurations, due to their identical mass number and comparable number of binary collisions ($N_{\text{coll}}$). Thus, the generic nature of the photon anisotropic flow would remain the same even after inclusion of the prompt calculation.

\section{Summary and conclusions}
Isobaric collisions at RHIC have recently attracted significant interest as they can be used for probing nuclear structure and for placing meaningful constraints on initial state models used to study the bulk properties of the matter created in such collisions. In this work, the initial state and subsequent evolution of the most central Ru+Ru and Zr+Zr collisions at 200A GeV at RHIC are investigated in detail using appropriate initial conditions and the MUSIC hydrodynamical model. Two extreme collision geometries are considered for both nuclei; the tip–tip configuration, which has the minimum initial spatial eccentricity and the body–body configuration, which exhibits the maximum spatial eccentricity.

The evolution of the average temperature is found to be largely insensitive to the collision orientation, with only a small difference observed at the earliest times. In contrast, the initial eccentricity of the overlap region has a more pronounced effect on the development of the average transverse flow velocity. The mildly prolate shape of ruthenium leads to a slower build-up of average transverse flow velocity compared to a circular shape or tip–tip collisions. For zirconium nuclei, the initial triangular anisotropy does not significantly alter the growth of transverse flow velocity compared to the tip–tip configuration.

The charged particle spectra and thermal photon spectra are found to be unaffected by the initial eccentricity of the collision zone, appearing similar for both isobaric systems. In contrast, the anisotropic flow coefficients are significantly larger for the body–body orientations in both Ru+Ru and Zr+Zr collisions at RHIC.

For zirconium, the larger value of initial triangular eccentricity ($\epsilon_3$) compared to the elliptic eccentricity ($\epsilon_2$) in ruthenium leads to a correspondingly larger triangular flow ($v_3$) than elliptic flow ($v_2$) for both charged particles and thermal photons.

The sensitivity of these results to the initial formation time and freezeout temperature indicates that photon anisotropic flow is considerably more sensitive to the initial state than hadronic observables. 

It has recently been shown that the ratio of the photon anisotropic flow parameter ($v_2/v_3$) along with the individual flow parameters can be a clean probe for investigating the initial state produced in relativistic heavy ion collisions~\cite{ratio, ratio1}. In addition, the ratio as a function of $p_T$ has been found to give a batter explanation of the experimental data compared to the individual flow coefficients. 

Therefore, direct photon measurements from isobaric collisions hold significant potential for isolating the effects of nuclear deformation from those of a spherical geometry as well as for constraining the properties of the initial state produced in relativistic heavy ion collisions.

A more realistic calculation, incorporating event-by-event fluctuating initial conditions and including the effects of viscosity, would provide deeper insights. We postpone such an investigation to a future study~\cite{amit}.

\begin{acknowledgments}
The authors would like to thank Jajati K. Nayak, Pingal Dasgupta, Sinjini Chandra, and Sanchari Thakur for their valuable discussions and support during the course of this study. A.P. also acknowledges the use of the Kanaad computing facilities at VECC.
\end{acknowledgments}

\end{document}